\documentclass[conference]{IEEEtran}

\usepackage{cite}
\usepackage{algorithmic}
\usepackage{algorithm}
\usepackage{array}
\usepackage{xcolor}
\usepackage[caption=false,font=normalsize,labelfont=sf,textfont=sf]{subfig}
\usepackage{textcomp}
\usepackage{stfloats}
\usepackage{subcaption}
\usepackage{url}
\usepackage{verbatim}
\usepackage{graphicx}
\usepackage{multirow}
\usepackage{amsmath,amssymb,amsfonts,amsthm}
\usepackage{algorithmic}
\usepackage{dsfont}
\usepackage{tikz}
\usetikzlibrary{shapes.geometric, arrows, arrows.meta, calc, shapes.misc, matrix, shapes, arrows, calc, positioning, shapes.geometric, shapes.symbols, shapes.misc, intersections, chains, scopes, fit}

\usepackage{optidef}
\usepackage{bm}
\usepackage{mathtools}
\usepackage[cmintegrals]{newtxmath}
\usepackage{glossaries-extra}
\usepackage{stackengine}
\usepackage{comment}
\usepackage[utf8x]{inputenc}

\newcommand{\calI}{\mathcal{I}}

\newcommand{\calS}{\mathcal{S}}
\newcommand{\calT}{\mathcal{T}}
\newcommand{\calU}{\mathcal{U}}

\newcommand{\red}[1]{\textcolor{red}{#1}}

\makeatletter
\newcommand\footnoteref[1]{\protected@xdef\@thefnmark{\ref{#1}}\@footnotemark}
\makeatother

\IEEEoverridecommandlockouts

\begin{document}

\title{Collective Grid: Privacy-Preserved Multi-Operator Energy Sharing Optimization via Federated Energy Prediction}

\author{
\IEEEauthorblockN{Meysam Masoudi\IEEEauthorrefmark{1},
Tahar Zanouda\IEEEauthorrefmark{1},
Milad Ganjalizadeh\IEEEauthorrefmark{1},
Cicek Cavdar\IEEEauthorrefmark{2}}

\IEEEauthorblockA{\IEEEauthorrefmark{1} Ericsson AB, Stockholm, Sweden\\
\texttt{\{meysam.masoudi, tahar.zanouda, milad.ganjalizadeh\}@ericsson.com}} \\
\IEEEauthorblockA{\IEEEauthorrefmark{2}KTH Royal Institute of Technology, Stockholm, Sweden\\
\texttt{\{cavdar\}@kth.se}}
}

\maketitle
\begin{abstract}
Electricity consumption in mobile networks is increasing with the continued 5G expansion, rising data traffic, and more complex infrastructures. However, energy management is often handled independently by each mobile network operator (MNO), leading to limited coordination and missed opportunities for collective efficiency gains. To address this gap, we propose a privacy-preserving framework for automated energy infrastructure sharing among co-located MNOs. Our framework consists of three modules: (i) a federated learning-based privacy-preserving site energy consumption forecasting module, (ii) an orchestration module in which a mixed-integer linear program is solved to schedule energy purchases from the grid, utilization of renewable sources, and shared battery charging or discharging, based on real-time prices, forecasts, and battery state, and (iii) an energy source selection module which handles the selection of cost-effective power sources and storage actions based on predicted demand across MNOs for the next control window. Using data from operational networks, our experiments confirm that the proposed solution substantially reduces operational costs and outperforms non-sharing baselines, with gains that increase as network density rises in 5G-and-beyond deployments.
\end{abstract}

\begin{IEEEkeywords}
Artificial intelligence for mobile networks, federated learning (FL), energy management, infrastructure sharing.
\end{IEEEkeywords}

\section{Introduction}\bstctlcite{IEEEexample:BSTcontrol}

With the evolution of mobile networks, growing traffic demands, and denser antenna deployments per site, network energy consumption continues to rise. To minimize operational expenditures (OPEX) and curb emissions, mobile network operators (MNOs) across the globe are developing innovative energy solutions and advanced orchestration strategies. Energy orchestration denotes the automated management, coordination, and optimization of energy production, storage, distribution, and consumption throughout the network. When effectively implemented, it empowers MNOs to significantly reduce costs while enhancing energy efficiency.
Advances in battery systems and energy-harvesting technologies, combined with energy orchestration, enable MNOs to capture and exploit small-scale local generation. This capability unlocks new business models and revenue streams, including cost savings via optimized energy management and strategic time-of-use interactions with the electricity grid~\cite{labidi2018optimal}.
Building upon this, incorporating renewable sources (e.g., solar and wind) is a promising strategy to reduce grid energy costs and carbon footprint~\cite{masoudi2019green}. However, renewable energy is weather-dependent and may not provide a consistently reliable source for mobile networks. As a result, MNOs must dynamically manage heterogeneous energy sources in real time, switching sources as demand, prices, and availability change. 
AI-driven orchestration offers a systematic path to build resilient, cost-efficient, and sustainable networks. Nevertheless, research has tended to focus on single MNO solutions. 
For example, \cite{labidi2018optimal} minimized one MNO’s electricity cost under smart-grid tariffs by optimizing battery charge/discharge, \cite{el2020battery} performed joint energy–radio scheduling for a single base station (BS) with hybrid grid supply while explicitly modeling battery degradation, and \cite{liu2021cooperative} developed an integrated planning framework for a 5G BS assisted by renewable resources and a battery-swapping system to minimize annualized cost. Despite progress in single MNO settings, cross MNO orchestration has received relatively limited attention.

Although improving energy use in isolation can be beneficial, such efforts fail to address energy challenges from a broader mobile network ecosystem perspective, particularly in multi-MNO settings. 
Deploying on-site multi-source energy infrastructure (i.e., harvesting, storage, and power electronics) typically entails substantial capital expenditure (CAPEX), which limits per-site viability. 
Infrastructure sharing offers a convincing alternative: by pooling investments, multiple MNOs can build a shared energy ecosystem that reduces costs and improves efficiency. This cooperative model, however, introduces new requirements, including accurate cross-MNO energy demand prediction and the need to strictly protect network data privacy.

To bridge this gap, recent industry initiatives in energy sharing~\cite{EricEnergySharingWhitePaper} introduce a paradigm shift toward shared renewable installations and backup storage across MNOs. By pooling resources, such approaches have the potential to reduce both CAPEX and OPEX while enhancing resilience through improved load balancing and geographic diversification of supply and demand.

In this paper, we propose a privacy-preserving framework for automated energy-infrastructure sharing among co-located MNOs. The framework is designed to minimize operational costs while improving end-to-end orchestration efficiency. Our key contributions are as follows:

\begin{itemize}
  \item  We design a federated learning (FL) assisted framework that coordinates energy-source selection for cross-MNO energy sharing while protecting network data privacy.
  
  \item  We formulate an optimization problem to minimize the MNOs’ operational costs. We compare the scenarios with and without shared energy infrastructure, considering renewable generation, grid prices, and storage dynamics.
  
  \item We evaluate our solution using operational network data and complementary simulations. Our results demonstrate cost-efficient power-supply strategies across a range of traffic loads, renewable availability, and storage capacity.
\end{itemize}

\section{System Model}

\begin{figure}
    \centering
    \includegraphics[width=0.75\linewidth]{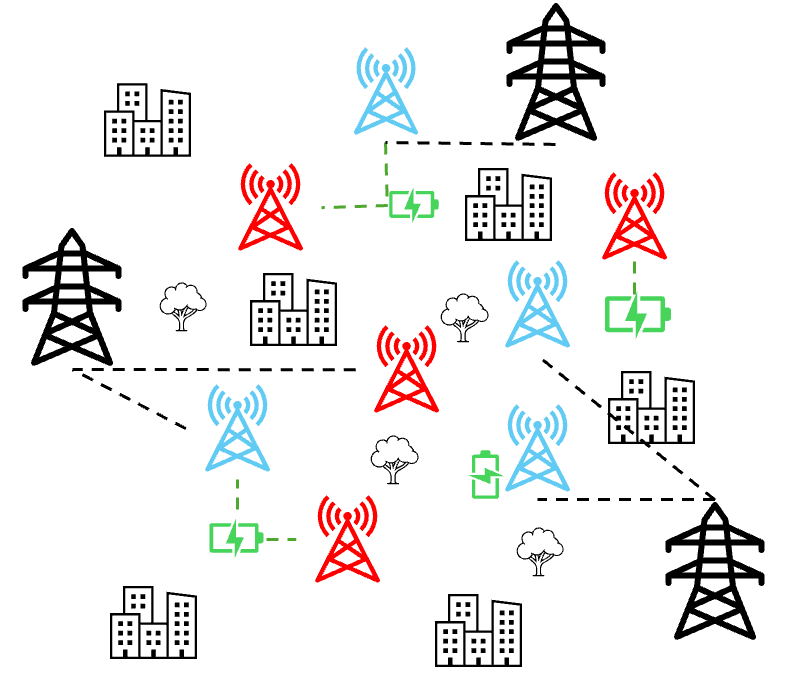}
    \caption{System model for a multi-MNO area in which sites are powered either by the grid or a shared battery storage. Sites belonging to different MNOs are distinguished by color.}
    \label{fig:sys}
\end{figure}

We consider a set of co-located MNOs $\mathcal{M}{\coloneqq}\{1,\ldots,M\}$ that operate radio sites $\mathcal{U}{\coloneqq}\{1,\ldots,U\}$. We define the radio site as a physical radio access network (RAN)  installation consisting of one or more BSs. Figure~\ref{fig:sys} illustrates the system in which the $M$ MNOs cover a given geographic area. Each site $u\in\mathcal{U}$ is owned by a unique MNO, captured by the mapping $\phi:\mathcal{U}{\to}\mathcal{M}$, where $\phi(u)$ denotes the MNO of site $u$. 
Energy is delivered via on-site energy infrastructures $\mathcal{I}{\coloneqq}\{1,\ldots,I\}$. Each site is served by a unique infrastructure specified by $\psi:\mathcal{U}{\to}\mathcal{I}$ and $i_u{\coloneqq}\psi(u)$. The set of sites fed by infrastructure $i$ is represented by $\mathcal{U}_i{\coloneqq}\{u\in\mathcal{U}:\psi(u)=i\}$.
Sites consume energy from a set of sources $\calS{\coloneqq}\{R,G,B\}$, where $R, G, B$ denote on-site renewables, electricity grid, and battery storage system, respectively. 
The battery capacity is defined at the infrastructure level as $\Omega_i$ in $\mathrm{kWh}$ for $i\in\mathcal{I}$.
Energy flows are modeled on a source-to-sink basis and are nonnegative.
The time is slotted into steps $\calT{\coloneqq}\{1,\ldots,T\}$ of duration $\Delta t$.
The grid and renewable energy to charge the battery at infrastructure $i$ is represented by $E^{G\to B}_{i}(t)$ and $E^{R\to B}_{i}(t)$, respectively. $E^{G\to L}_{i_u}(t)$ and $E^{B\to L}_{i_{u}}(t)$ denote the grid and battery energy from infrastructure $i_u$ (the infrastructure that serves $u$) to serve site $u$ for $u{\in}\calU$. 
Let $D_u(t)$ denote the demand of site $u$ during period $[t,t{+}\Delta t)$, and let $d_u(t){\in}\{0,1\}$ select the supply source at time $t$ ($d_u(t){=}0$ grid, $d_u(t){=}1$ battery). The per-period supply are then
\begin{equation}
E^{G\to L}_{i_u}(t)=(1-d_u(t))\,D_u(t),
\end{equation}
and
\begin{equation}
E^{B\to L}_{i_u}(t)=d_u(t)\,D_u(t).
\end{equation}

\subsection{Energy Sources}
\subsubsection{Renewable Energy}
Renewable energy generation depends on environmental conditions. Consequently, weather variability directly affects harvested energy. We collect site-specific meteorological features, obtained from third-party services or using on-site sensors, and the base-station geolocation (latitude, longitude, altitude). Typical features include solar irradiance, ambient temperature, humidity, cloud cover, and precipitation. Harvested sources interface with the on-site battery system and may include: (i) solar panels, (ii) wind turbines, and (iii) micro-hydro (water) turbines. In this study, we consider only solar energy, while the framework is not limited to it.

\subsubsection{Electrical Grid}
The system monitors electricity price signals to schedule purchases from and to the grid. Price data (e.g., day-ahead or real-time tariffs) is typically available via provider portals and can be accessed via  APIs \cite{ElectricityMaps}. These signals, combined with renewable forecasts and the battery state, are supplied to the optimizer to determine cost-effective import decisions.

\subsection{Battery States Model}
The battery can power the BS and store energy for later use.
Each time we utilize the battery, two main characteristics of the battery change: (1) state of charge (SoC), which reflects the instantaneous charge level, and (2) state of health (SoH), which captures long-term degradation characteristics. 

The battery SoC at $i$th infrastructure and time $t$, $x_i(t)$(${\in}[0,1]$), is defined as the ratio of stored energy to the effective battery capacity. It evolves from the previous period by accounting for net charged and discharged energy, normalized by capacity: 
\begin{equation}
x_i(t)
= x_i(t-1){+}\frac{1}{\Omega_i}\left(E^{B,\mathrm{mix}}_i(t){-}\sum_{u\in\mathcal{U}_i} E^{B\to L}_{i_u}(t)\right).
\end{equation}
where the effective weighted charge input is
\begin{equation}
E^{B,\mathrm{mix}}_{i}(t)\coloneqq \lambda_i(t)\,E^{G\to B}_{i}(t) + \big(1-\lambda_i(t)\big)\,E^{R\to B}_{i}(t),
\end{equation}
where $\lambda_i(t){\in}[0,1]$. Here, $E^{G\to B}_i(t)$ and $E^{R\to B}_i(t)$ denote the energy used to charge the battery from the grid and from on-site renewables, respectively, and $E^{B\to L}_{i_u}(t)$ denotes the battery discharge serving the load.
Moreover, we use $\lambda_i(t){\in}[0,1]$ to model the charging mix, i.e., $\lambda_i(t){=}1$ and $\lambda_i(t){=}0$ represent grid-only and renewable-only, respectively, and $\lambda_i(t){\in}(0,1)$ is a convex combination of the two. 

The battery SoH at $i$th infrastructure and time $t$, $y_i(t)$, models capacity degradation resulting from both usage cycles and aging over time.
Following \cite{labidi2018optimal}, $y_i(t)$ can be derived as:
\begin{equation}
y_i(t)=y_i(t-1)-\kappa\,\big|\, x_i(t)-x_i(t-1)\big|,
\end{equation}
where $\kappa$ is a constant value from the battery configuration. One may define $\kappa$ as an increasing function in time and temperature to reflect aging and incorporate the impact of temperature on SoH. It is worth noting that the final SoH of the battery is obtained when the impact of all individual site u is considered on the SoH.
The battery capacity is then taken to be proportional to the SoH, as
\begin{equation}
\Omega_i \coloneqq \Omega_i^{\mathrm{Init}}\,y_i(t),
\end{equation}
where $\Omega_i^{\mathrm{Init}}$ is the battery capacity at $t=0$.

\subsection{Energy Infrastructure Total Cost}
The total cost of ownership, $\Gamma^{\mathrm{Total}}_{u}$, to provide sites' energy is composed of CAPEX and OPEX which is calculated as:
\begin{align}
\Gamma^{\mathrm{Total}}_{u}&\coloneqq 
\underbrace{\alpha_{i_u}\left(\Gamma^{\mathrm{Infra}}_{i_u}+\Gamma^{\mathrm{Battery}}_{i_u}+\Gamma^{\mathrm{Renew}}_{i_u}\right)}_{\text{CAPEX of $u$ from infrastructure }i}+
\notag\\
&\underbrace{\alpha_{i_u}\Gamma^{\mathrm{Rent}}_{i_u} + \sum_{t\in\mathcal{T}}\left(p^{G}\!(t)\,\left(E^{G\to L}_{i_u}\!(t)+\alpha_{i_u}\,E^{G\to B}_{i}\!(t)\right)
\right)}_{\text{OPEX}~u},
\end{align}
where $\Gamma^{\mathrm{Infra}}_{i_u}$, $\Gamma^{\mathrm{Battery}}_{i_u}$, and $\Gamma^{\mathrm{Renew}}_{i_u}$ are the cost of rolling out the power infrastructure, the battery cost,  and the cost of renewable energy infrastructure, respectively. Moreover,
$p^{G}(t)$ is the average electricity price during the $t$th time step, $\alpha_{i_u} \in (0,1]$ represents the share of infrastructure $i_u$ attributed to site $u$ (and hence the MNO $\phi(u)$). 
Shares are normalized per infrastructure, i.e., $\sum_{u\in\mathcal{U}_{i}} \alpha_{i_u}$ for all $i{\in}\calI$.
When the infrastructure is not shared with others, $\alpha_{i_u} {=} 1$. 
MNOs can decide on the policy to set $\alpha_{i_u}$, for example, equal share for sites (i.e., proportional to the number of sites owned by MNOs using this infrastructure) $\alpha_{i_u} = \frac{1}{|\calU_i|}$.  Total cost per network can be calculated as: 
\begin{equation}
    \Gamma^{\mathrm{Total}} = \sum_{u \in \calU} \Gamma^{\mathrm{Total}}_{u},  
\end{equation}
For our evaluations, we define $\zeta$ as the ratio of radio sites sharing the infrastructure as a design parameter, formulated as
\begin{equation}
\zeta =  \frac{\sum_{u\in\calU} \mathds{1}{\{\alpha_{i_u}=1\}}}{U}
\end{equation}
where $\mathds{1}{\{\cdot\}}$ is $1$ when the condition $\alpha_{i_u}=1$ holds.

\section{Energy Infrastructure Sharing Framework }

 \begin{figure}[!t]
\centering
  \includegraphics[width=.89\linewidth]{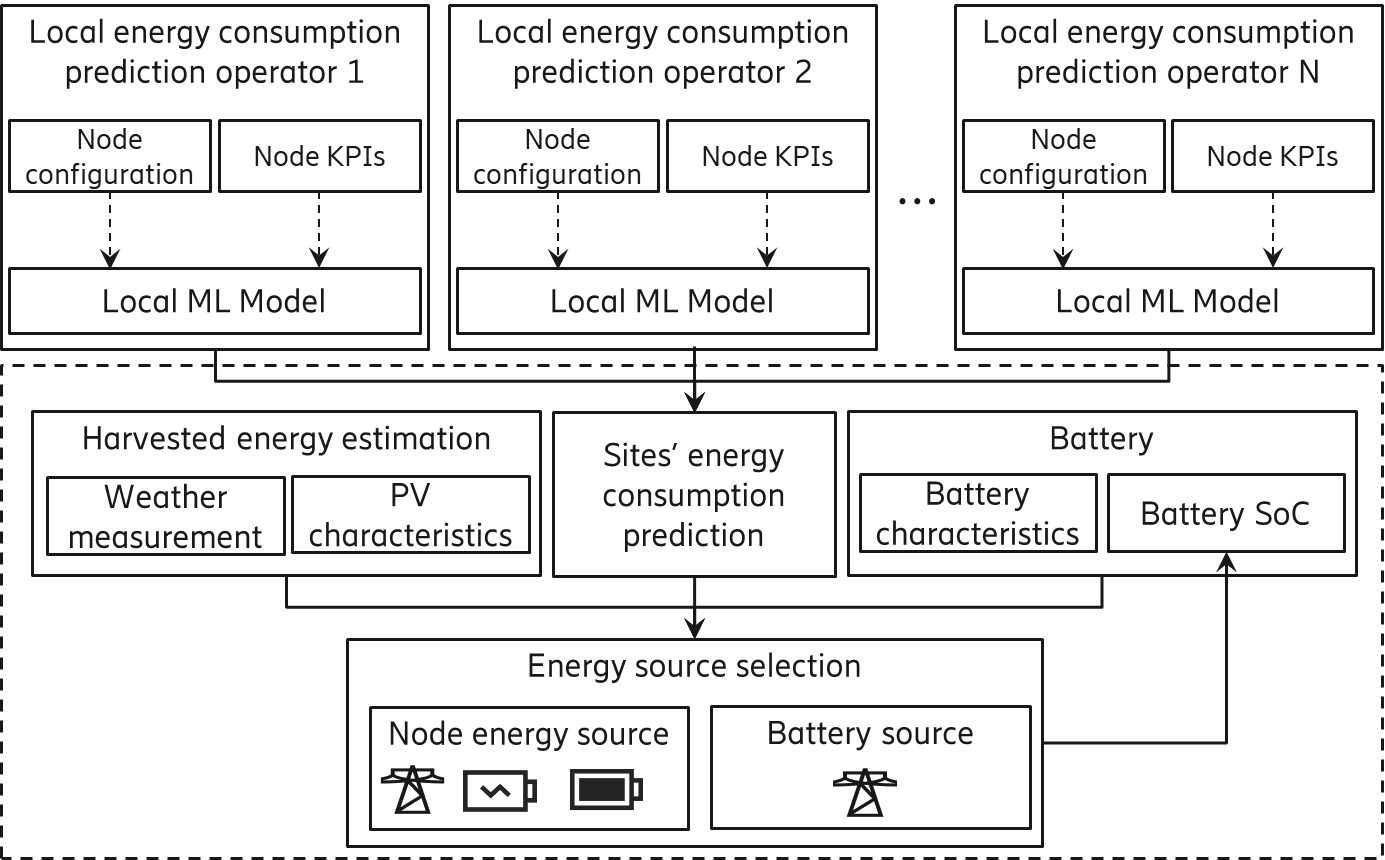}
  \caption{Multi-MNO power infrastructure sharing and battery management framework}
  \label{fig:main_method}
\end{figure}%

In this section, we explain our proposed privacy-preserving energy infrastructure sharing framework, depicted in  Fig.~\ref{fig:main_method}. 
\subsection{Energy Consumption Prediction}

\subsubsection{RAN Data}
\label{randata}
Mobile network data primarily consists of time-series data and system logs, capturing the dynamic behavior of network elements. Mobile network data can be broadly categorized into two types: static and dynamic. {Static data} includes information about the mobile network infrastructure, such as power equipment, data centers, inventory warehouses, which is typically stored in the {Mobile network infrastructure inventory and configuration} databases. This data changes infrequently and represents the foundational structure of the network. {Dynamic data}, on the other hand, captures the ongoing behavior of the network in operation. This includes:
\begin{itemize}
    \item \textbf{Configuration Management (CM) data:} Represents the structural characteristics of the network through various configuration parameters (e.g., frequency bands, technology type, hardware models).
    \item \textbf{Performance Management (PM) data:} Collected regularly across various software and hardware components of the network. PM data tracks operational metrics that reflect real-time performance.
\end{itemize}

To evaluate network performance, MNOs rely on {Key Performance Indicators (KPIs)} which are standardized metrics derived from CM and PM data. These KPIs are grouped into several categories, including: accessibility (e.g., successful connection attempts, cell data rates), mobility, utilization, energy performance, etc. In this study we take into account
{accessibility KPIs} (such as successful connections and throughput metrics) and {node and power configuration parameters} from CM data. These are used to model and estimate energy usage patterns across the RAN.

\subsubsection{Federated Learning}

FL enables multiple MNOs to collaboratively train predictive models for energy demand without exchanging raw data, thereby preserving privacy and security \cite{Konecny2016,McMahan2016}. In mobile networks, each site  trains a local time–series model on its own measurements (traffic, weather, renewable energy, battery state) and periodically uploads model updates to a coordinator, which aggregates them into a global model (e.g., via FedAvg~\cite{McMahan2016}) for improved forecasting and energy orchestration. This paradigm aligns well with dynamic, geographically distributed networks where data residency and confidentiality are stringent. 
While FL has proven practical, several challenges remain in our setting:
i) the non-IID input data across sites and MNOs can cause client drift and slow or destabilize convergence,
ii) system constraints (e.g., unreliable backhaul, stragglers) increase communication overhead, and
iii) concept drift driven by traffic seasonality or weather conditions can make global models outdated frequently
One of the approaches to mitigate these challenges is through weighted aggregation and periodic retraining~\cite{Konecny2016,McMahan2016}. Under these design choices, FL remains a viable solution for privacy-preserving energy consumption prediction for multi-MNO networks.

\subsubsection{FL-based Power Consumption Prediction}

We predict site-level energy consumption for the next control window $t{+}\Delta$ using a multivariate long short-term memory (LSTM), trained continuously in a federated setup to preserve data privacy \cite{hochreiter1997long}. To capture temporal and cross-operator variability, the input for each site $u$ is a sliding window of KPIs over the last four hours, throughput and successful connection ratio sampled every 15 minutes (16 samples per feature), augmented with CM data that is static over the window. Specifically, the per-site feature vector comprises:
(i) KPI sequences $\{\,\text{throughput}_u(\tau),\ \text{success\_ratio}_u(\tau)\,\}_{\tau=t{-}15, \ldots, t}$ and
(ii) CM data (hardware model, power-configuration parameters, and frequency band).
The LSTM encodes the KPI sequence while the CM data are concatenated at the appropriate layer to condition the prediction. The model outputs $\widehat{D}_u(t{+}\Delta)$, the predicted energy consumption demand for site $u$. Federated training aggregates sites' updates with weighted averaging, enabling continual adaptation across heterogeneous MNOs without centralizing raw data.
The prediction procedure is:

\begin{enumerate}
    \item \textbf{Site selection:} At each round at time $t$, a scheduler selects available RAN nodes within each installation site according to a selection rule (e.g., availability). 
    \item \textbf{Parameter broadcasting:} The central server broadcasts the current global weights $\omega_t$ to the selected sites.
    \item \textbf{Local model training:} Each site trains the LSTM locally on its collected dataset  to minimize a prediction loss (e.g., MSE), producing updated weights $\omega_t^{(i)}$.
    \item \textbf{Upload and aggregation:} Clients upload their local updates to the server, which aggregates them to form a new global model. The updated model is then used to produce estimates of per-site and total energy consumption.
    \item \textbf{Convergence:} Steps 1–4 repeat until a stopping criterion is met (e.g., validation loss, maximum rounds).
\end{enumerate}

We use the vanilla FedAvg rule \cite{McMahan2016} to aggregate local models:
\begin{equation}
    \omega_{t+1} \;=\; \sum_{u \in \calS} \frac{n_u}{\sum_{i \in \calS} n_i}\,\omega_t^{(u)},
    \label{eq:FedAvg}
\end{equation}
\noindent where $\calS$ is the set of selected sites at round $t$, $n_u$ is the number of local samples at site $u$, each $\omega_t^{(u)} {\in} \mathbb{R}^{N_{LM}}$ denotes a site’s model-parameter vector, and the global weights $\omega_{t+1} $ are broadcast for the next round.

\subsection{Energy Cost Optimization Problem} \label{sec:opt}

The total TCO minimization problem at each decision time $t$ and $\forall u\in\mathcal{U}, \forall i\in\mathcal{I}, \forall t\in\mathcal{T}$ can be formulated as:

\begin{subequations}\label{eq:objective}
\begin{alignat}{3}
\min_{\{d_u(t)\}_{u\in\calU}} \;\; & \sum_{u\in\calU} \Gamma^{\mathrm{Total}}_{u} \tag{\ref{eq:objective}}\\
\text{s.t.} \quad 
& x_{i}(t) \ge SoC^{\min}, \quad\forall i\in\mathcal{I}\label{eq:sOC}\\[1mm]
& y_i(t) \geq SoH^{\min}, \quad\forall i\in\mathcal{I}\label{eq:sOH}\\[1mm]
& \sum_{u\in \calU{i}} E^{B\to L}_{i_u}(t)  \leq \Omega_i\, \mathds{1}{\left\{\sum_{u{\in}\calU_i} d_u(t)> 0\right\}} , \label{eq:battery_cap}\\[1mm]
& d_u(t) \in \{0, 1\}.
\end{alignat}
\end{subequations}

\noindent
The objective function minimizes the sum of the total costs $\Gamma^{\mathrm{Total}}_{u}$ over all sites using the decision variable $d_u(t)$ which select grid, $d_u(t){=}0$, or battery, $d_u(t){=}1$, to serve the site $u$.  Constraint \eqref{eq:sOC} ensures that the state-of-charge for each battery is above a minimum threshold $SoC^{\min}$. Constraint \eqref{eq:sOH} ensures that the state-of-health for each battery is above a minimum threshold $SoH^{\min}$. Constraint \eqref{eq:battery_cap} ensures that if we use battery energy (i.e., $d_u = 1$), then the total energy consumption $D_u$ does not exceed the battery capacity $\Omega_u$. 

The energy source for the next time window (i.e., $t{+}\Delta t$) is selected by solving the optimization problem. The inputs are live electricity prices from external markets, the predicted energy consumption, the forecast of harvested (renewable) energy, and the current battery state (e.g., state of charge and availability). The problem is formulated as a mixed-integer linear program (MILP) and is solved using existing optimization solvers, i.e., milp in SciPy. The prediction/control window, i.e., $\Delta t$, is chosen by the energy infrastructure but is lower-bounded by the granularity of KPI collection. Short windows can induce a \emph{ping-pong} effect (frequent source switching), whereas longer windows increase the risk of battery depletion due to forecast error and may miss cost-saving opportunities when prices or loads are low. 
\vspace{-10pt}
\subsection{Infrastructure Sharing Framework}
The proposed sharing method comprises five main stages:
\subsubsection{Data Collection}  At this stage, each site collects its own KPIs from CM and PM data.  

\subsubsection{FL-based Power Consumption Prediction} This stage has two main steps: i) local model training ii) global model prediction.
In the first step, each site trains its own local model for energy consumption prediction and fine tune its model weights based on the recent collected data, and previously shared weights from the global model. Then, it shares the model weights  to the coordinator. 
In the second step, the coordinator collects all the information from the sites and aggregate them. At this stage, the coordinator predicts the energy consumption of each site and estimates the total energy consumption together with  harvested energy. 

\subsubsection{Decision-making} The estimated energy consumption and harvested energy values are sent to a decision-making entity. This entity considers various factors, including the energy required for the next time step, battery status, safety margins, and other operational constraints, to determine the optimal power supply strategy. The decision involves selecting the energy source for the subsequent time steps, which could be from the grid, harvested energy, or neither, for energy storage purposes. To make the decision, we solve the optimization problem explained in Section~\ref{sec:opt}.  We can solve this problem at the required granularity of at least every 15 minutes. 

\subsubsection{Energy Source Switching} The system switches the energy source to the selected option and powers all sites based on their energy demands. 

\subsubsection{Monitoring} At this stage, the framework monitors the actual and predicted energy consumption, KPIs sent by the MNOs, and other conditions such as weather, to make sure that the energy source remains optimal for the next time window. 
For instance, suppose that the battery is selected as the power source for the next time window. If the required energy exceeds the available battery capacity (including a marginal safeguard), the framework may override the current decision and trigger a manual switch of the power source to the grid until the next decision time. In this way, the framework ensures that unintended power outages at the sites do not occur.

\section{Simulation Results}

We collected data from a European region in which three MNOs cover the same area. There are a total of $1200$ sites, and we assumed all sites are connected to the grid, with a subset equipped with backup batteries. Each site may have multiple cells, ranging from 3 to 9 cells, and all cells employ $64$-antenna MIMO. The power consumption of sites includes cooling, transmission, processing, etc., of all cells on the same site. The dataset includes 7 days of data, with a time granularity of 15 minutes, and has information about CM and PM data explained in Section~\ref{randata}.
The simulations are performed in Python using Pytorch and FL Framework. The simulation parameters are summarized in Table \ref{tab:cost}.
We fine-tuned the LSTM-FL model to improve the prediction performance. Table \ref{tab:mae_results_fl} summarizes the mean square error (MSE) and mean absolute error (MAE) for energy prediction. The best result is obtained with a sequence length of $16$, $10$ iterations, and $3$ training epochs per round. 

\begin{table}[]
\footnotesize
\caption{Simulation parameters.}
\label{tab:cost}
\renewcommand{\arraystretch}{1.3}
\begin{tabular}{|l|l|l|l|l|l|}
\hline
Par. & Value  & Par.  & Value& Par.  & Value\\
\hline
$\Omega_u$ & $50$\,kWh& $P^{G}$ & $0.28$\,€/kWh & $\Gamma^{\mathrm{Renew}}_u$ & $3$\,k€\\
$\Gamma^{\mathrm{Rent}}_u$& $50$\,k€/year&$\Gamma^{\mathrm{Battery}}_u$&10\,k€&$SoC^{\min}$&0.1\\
$\kappa$& $8.85e^{-6}$&$T$/$\Delta t$& 15\,years/15\,min&$U$&1200\\
\hline
\end{tabular}
\end{table}

\begin{table}[]
\footnotesize
\caption{MSE and MAE for LSTM using FL.}
\label{tab:mae_results_fl}
\renewcommand{\arraystretch}{1.1}
\begin{tabular}{|l|l l l|l l l|}
\hline
Params &   & MSE  &   &   & MAE & \\
\hline
[seq len, iter, epoch] & Min & Mean & Max & Min & Mean & Max\\\hline
$[ 16, 10, 3]$ & 0.13 & 0.51 & 1.11 & 0.27 & 0.39 & 0.54  \\
$[16, 50, 10]$ & 0.18 &  0.58 & 1.18 & 0.31 & 0.43 & 0.58  \\ 
$[32, 10, 3]$ & 0.13 &  0.51 & 1.12& 0.27 & 0.39 & 0.53  \\ 
$[32, 50, 10]$ & 0.19 &  0.58 & 1.19 & 0.30 & 0.43 & 0.58  \\ 
$[64, 10, 3]$ & 0.13 &  0.51 & 1.12& 0.27 & 0.39 & 0.53  \\
$[64, 50, 10]$ & 0.18 &  0.59 & 1.18 & 0.32 & 0.43 & 0.58  \\ 
\hline
\end{tabular}
\end{table}

\begin{figure}[!t]
    \centering
\includegraphics[trim={1.5cm .7cm 1cm 1cm},clip,width=0.98\linewidth]{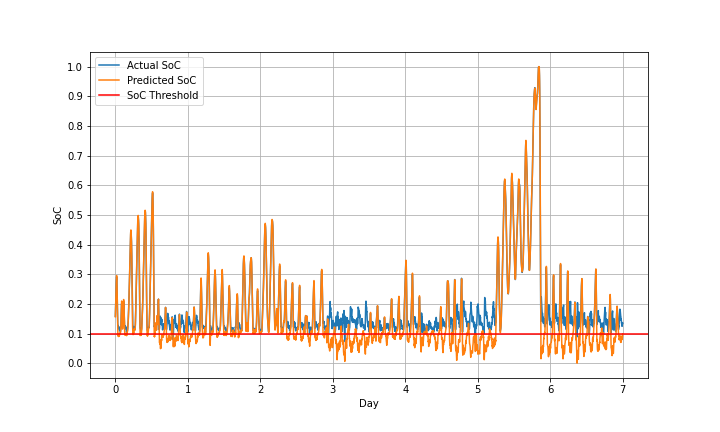}
    \caption{Predicted and actual SoC based on the predicted and actual power consumption.}
    \label{fig:SoC}
\end{figure}

Fig. \ref{fig:SoC} shows the SoC for a site with six cells. If the SoC falls below $10\%$ of battery capacity, only grid electricity is used, ensuring a minimum battery level for emergencies like power outages or inaccurate power predictions. The figure demonstrates that power predictions are accurate for forecasting SoC. Predicted SoC values below the threshold indicate times when predicted power consumption could drain the battery beyond this limit, necessitating the use of the primary power source to prevent further decline in SoC.

\begin{figure}[!t]
    \centering    \includegraphics[trim={0cm .0cm .0cm 0cm},clip,width=0.98\linewidth]{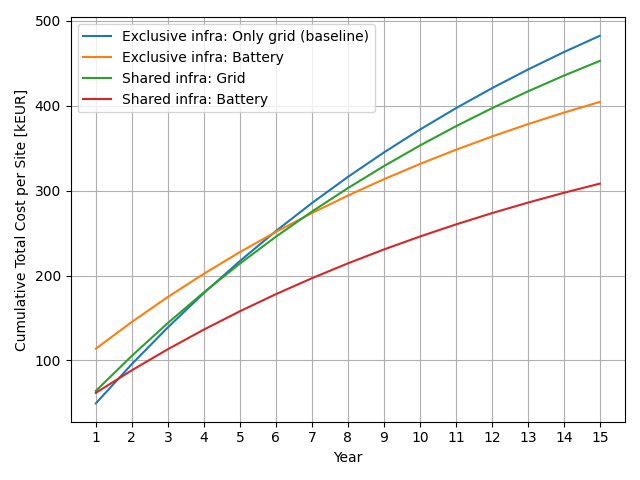}
    \caption{Cumulative total cost per site over a 15-year period for different infrastructure and battery configurations. }
    \label{fig:SiteCost}
    \vspace{-5mm}
\end{figure}

\begin{figure}[!t]
    \centering
    \includegraphics[trim={0.25cm .05cm 0cm .85cm},clip,width=0.98\linewidth]{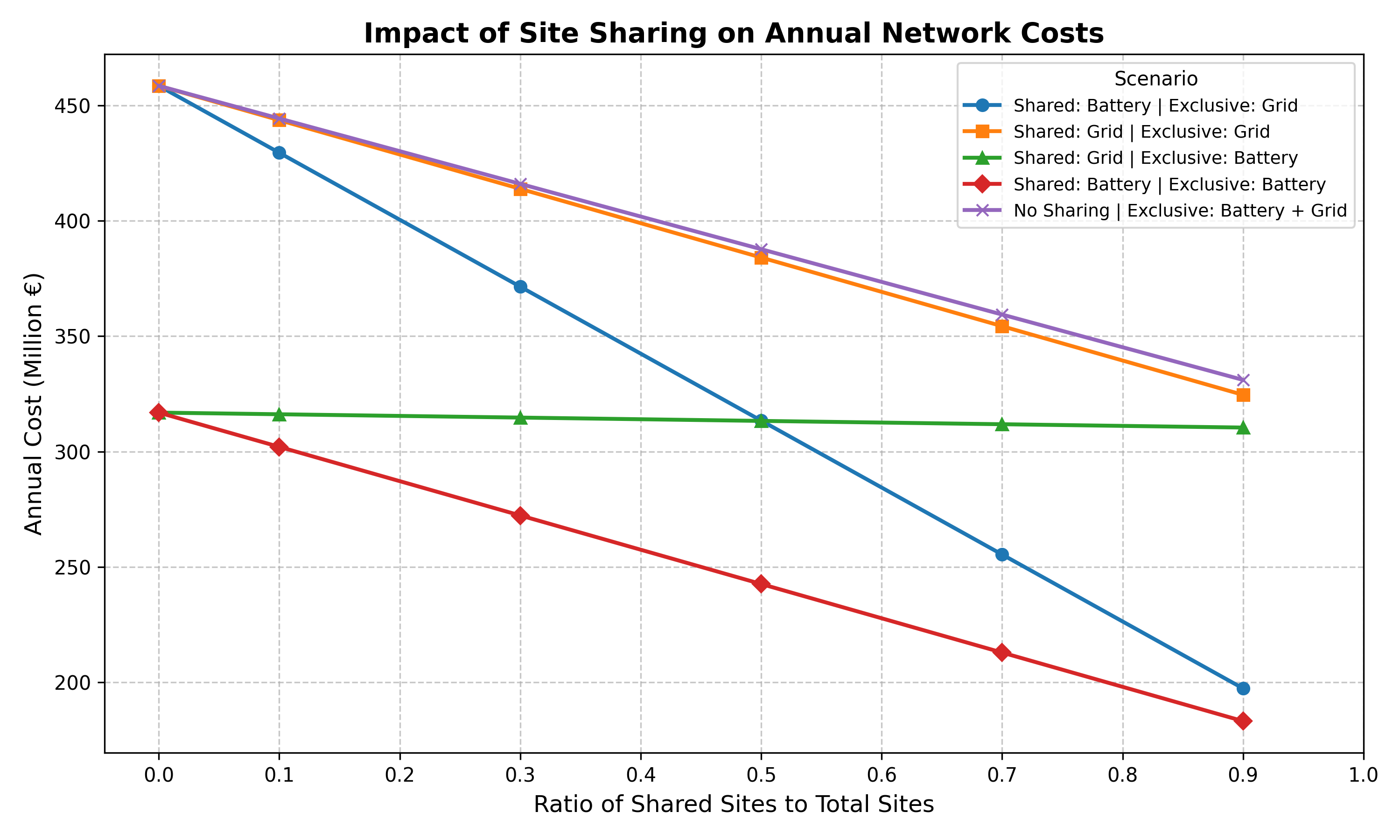} \caption{Annual operational cost as a function of $\zeta$, i.e., the site-sharing ratio. }
    \label{fig:annual}
    \vspace{-5mm}
\end{figure}

Figure \ref{fig:SiteCost} illustrates the cumulative total cost per site over a 15-year period for four configurations of power infrastructure and battery integration. The baseline represents exclusive infrastructure powered only by the grid, where each MNO independently deploys and operates its sites. Adding batteries in the non-sharing setup moderately reduces total costs, as the higher initial CAPEX is offset by lower grid-energy usage over time. Sharing infrastructure without batteries yields further savings through distributed investment and operational costs. The most cost-effective configuration occurs when MNOs share both the infrastructure and a common battery system, achieving the lowest cumulative cost across the analysis period. The crossover points mark a shift from CAPEX- to OPEX-dominated cost dynamics, while grid-only sharing minimizes early expenditures. Battery-based configurations, especially with shared storage, show the best long-term performance.
Fig.~\ref{fig:annual} illustrates the annual cost of a network with $1200$ sites. In this figure, a shared battery configuration implies that both the battery and the grid connection are shared. The site-sharing ratio $\zeta$ is defined as the fraction of sites that participate in sharing, i.e., the number of shared sites divided by the total number of sites.
We consider five scenarios. The first four are parameterized by $\zeta$ and described as
Shared: \{Battery or Grid\} $\mid$ Exclusive: \{Battery or Grid\}:
the “Shared” part specifies the technology used at the $\zeta$ fraction of shared sites, and the “Exclusive” part specifies the technology used at the remaining $(1-\zeta)$ fraction of non-shared sites. In this notation, “Battery’’ means that sites use both a battery and the grid, whereas “Grid’’ means a grid-only supply.
The fifth benchmark (purple) corresponds to a fully non-shared setting in which all sites are exclusive. In this case, $\zeta$ is used as the fraction of sites equipped with dedicated batteries, with the remaining $(1-\zeta)$ sites being grid-only.
This figure demonstrates that increasing the degree of infrastructure sharing among MNOs leads to substantial cost savings. Moreover, increasing the sharing ratio reduces yearly costs, with sharing always being cheaper. 
The results further indicate that it is advantageous to equip most sites with batteries. For instance, if most sites are exclusive, sharing only the grid while equipping exclusive sites with batteries is beneficial, as the cost reduction from shared sites is less than the CAPEX, and exclusive sites gain more from utilizing batteries.

\begin{figure}[!t]
    \centering
    \includegraphics[width=0.85\linewidth]{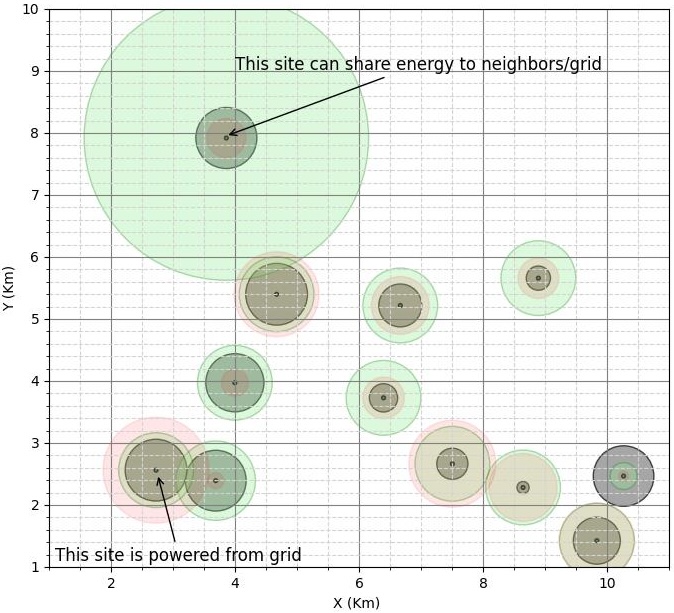}
    \caption{Energy storage over time and location. Green circles show the battery charge, red circles show the required energy. }
    \label{fig:geo}
    \vspace{-5mm}
\end{figure}

Fig. \ref{fig:geo} illustrates how the battery charge changes over time and location. The green circles represent the current battery charge, the red circles show the required energy for the next time step, and the black circles show the previous battery charge. With the proposed approach, energy can be stored in different geographical locations and hence can be used for either selling to the grid or sharing energy with neighbors.


\section{Conclusion}

In this paper, we presented our framework for energy orchestration with shared infrastructure across co-located MNOs. The framework forecasts per-site energy consumption via FL (preserving data privacy) and jointly optimizes  grid purchase schedules, renewable utilization, and shared battery charge/discharge to supply RAN nodes sustainably and cost-effectively. Using data from operational networks, our evaluation confirmed that the proposed sharing approach reduces operational expenditures compared to non-sharing baselines. We also quantified the costs of migrating to a shared energy infrastructure and identified a crossover point beyond which the higher upfront investment in renewables and storage yields in OPEX savings. The benefits increase with network densification and greater storage/renewable penetration. 
Our results provide valuable insights into cross-MNO infrastructure sharing, demonstrating how such joint energy scheduling convert higher upfront investment into long-term OPEX reductions.

\bibliographystyle{IEEEtran}
\bibliography{IEEEabrv,bibliography.bib}

@IEEEtranBSTCTL{IEEEexample:BSTcontrol,
CTLuse_article_number = "yes",
CTLuse_paper = "yes",
CTLuse_url = "yes",
CTLuse_forced_etal = "yes",
CTLmax_names_forced_etal = "1",
CTLnames_show_etal = "1",
CTLuse_alt_spacing = "yes",
CTLalt_stretch_factor = "1",
CTLdash_repeated_names = "yes",
CTLname_format_string = "",
CTLname_latex_cmd = "",
CTLname_url_prefix = "[Online]. Available:"
}

@inproceedings{McMahan2016,
  title={Communication-efficient learning of deep networks from decentralized data},
  author={McMahan, Brendan and Moore, Eider and Ramage, Daniel and Hampson, Seth and y Arcas, Blaise Aguera},
  booktitle={Artificial intelligence and statistics (AISTATS)},
  pages={1273--1282},
  year={2017}
}

@ARTICLE{Konecny2016,
  title = {Federated Optimization: Distributed Machine Learning for On-Device Intelligence},
  author={Jakub Konečný and H. Brendan McMahan and Daniel Ramage and Peter Richtárik},
  journal={\textup{2016}, 	arXiv:1610.02527 \textup{[cs.LG]}}
  }

@TECHREPORT{EricEnergySharingWhitePaper,
  author = {Anette Hoglund and et al},
  title = {Innovative Energy sharing},
  url = {https://www.ericsson.com/en/reports-and-papers/white-papers/innovative-energy-sharing},
  year={2024},
  INSTITUTION={Ericsson}
}

@article{masoudi2019green,
  title={Green Mobile Networks for {5G} and Beyond},
  author={Masoudi, Meysam and Khafagy, Mohammad Galal and Conte, Alberto and El-Amine, Ali and Fran{\c{c}}oise, Brian and Nadjahi, Chayan and Salem, Fatma Ezzahra and Labidi, Wael and S{\"u}ral, Altu{\u{g}} and Gati, Azeddine and others},
  journal={IEEE Access},
  volume={7},
  pages={107270--107299},
  year={2019},
  publisher={IEEE}
}

@article{hochreiter1997long,
  title={Long Short-term Memory},
  author={Hochreiter, S},
  journal={Neural Computation MIT-Press},
  year={1997}
}

@article{labidi2018optimal,
  title={Optimal battery management strategies in mobile networks powered by a smart grid},
  author={Labidi, Wael and Chahed, Tijani and Elayoubi, Salah-Eddine},
  journal={IEEE Transactions on Green Communications and Networking},
  volume={2},
  number={3},
  pages={859--867},
  year={2018},
  publisher={IEEE}
}

@online{ElectricityMaps,
  title = {Electricity Maps},
  url = {https://app.electricitymaps.com/map},
  urldate = {2023-03-14}
}

@article{liu2021cooperative,
  title={Cooperative planning of distributed renewable energy assisted 5G base station with battery swapping system},
  author={Liu, Xiyuan and Bie, Zhaohong},
  journal={IEEE Access},
  volume={9},
  pages={119353--119366},
  year={2021},
  publisher={IEEE}
}

@article{el2020battery,
  title={Battery-aware green cellular networks fed by smart grid and renewable energy},
  author={El Amine, Ali and Hassan, Hussein Al Haj and Nuaymi, Loutfi},
  journal={IEEE Transactions on Network and Service Management},
  volume={18},
  number={2},
  pages={2181--2192},
  year={2021},
  publisher={IEEE}
}

\end{document}